\documentclass[preprint,aps]{revtex4}
\usepackage{euscript}
\usepackage{amsmath}
\usepackage{epsfig}
\begin{document}
\preprint{UCI-TR-2008-35}
\title{Addition of First Generation leptons to the External Flux Model}
\author{Aaron Roy\footnote{Electronic address:roya@uci.edu}
}
\affiliation{
Department of Physics and Astronomy, University of California, Irvine,
California 92697-4575}

\begin{abstract}
In an extra dimensional EW model in M$_4 \bigotimes \, $S$_1$ there is no distinction mathematically with the standard model analog as far as the degrees of freedom of the two models along with the masses and more importantly the mass ratio relation in the zero mode limit. In this paper we present a theoretical construct of the same geometry but with the addition of an external magnetic flux permeating the extra coordinate. This will give all of the charged fields in the model an additional phase with nontrivial periodicity. This rather important addition leads to very interesting and mathematically rich physics. Here we will present the generalized theory for the addition of first generation leptons to this theory.      
\end{abstract}

\maketitle

\section{Addition of fermions to the model}
The following fermions are added to the model, $E_L = \begin{pmatrix} \nu ^e _L \\ e_L\end{pmatrix}$ for the left handed doublet electron and neutrino, $e_R$ for the right handed electron singlet, $Q_L = \begin{pmatrix} u_L \\ d_L\end{pmatrix}$ for the left handed up and down quark doublet, $u_R$ and $d_R$ for the right handed up and down quark singlets. Finally $E_R = \begin{pmatrix} \nu ^e _R \\ e_R \end{pmatrix}$ for the right handed electron and neutrino doublet and \\ $Q_R = \begin{pmatrix} u_R \\ d_R\end{pmatrix}$ for the right handed up and down quark doublet. The terms that we will add to the 5-D lagrangian density are \\
$\EuScript{L}_{\rm{fermions}} = i\begin{pmatrix} \bar{\nu}^e _L , & \bar{e}_L \end{pmatrix}D_\mu \gamma ^\mu \begin{pmatrix} \nu ^e _L \\ e_L \end{pmatrix} + i\bar{e}_R D_\mu \gamma ^\mu e_R + i\begin{pmatrix} \bar{u}_L , & \bar{d}_L \end{pmatrix}D_\mu \gamma ^\mu \begin{pmatrix} u_L \\ d_L \end{pmatrix} \\ + i\bar{u}_R D_\mu \gamma ^\mu u_R +  i\bar{d}_R D_\mu \gamma ^\mu d_R - \begin{pmatrix} \bar{\nu}^e _L , & \bar{e}_L \end{pmatrix}D_5 \gamma ^5 \begin{pmatrix} \nu ^e _R \\ e_R \end{pmatrix} - \begin{pmatrix} \bar{u}_L , & \bar{d}_L \end{pmatrix}D_5 \gamma ^5 \begin{pmatrix} u_R \\ d_R \end{pmatrix} \\ - \lambda _e \begin{pmatrix} \bar{\nu}^e _L , & \bar{e}_L \end{pmatrix}\phi e_R - \lambda _d \begin{pmatrix} \bar{u}_L , & \bar{d}_L \end{pmatrix}\phi d_R - \lambda _u \begin{pmatrix} \bar{u}_L , & \bar{d}_L \end{pmatrix}i\tau ^2 \phi ^* u_R$ where $i\tau ^2 = \begin{pmatrix} 0 & 1 \\ -1 & 0 \end{pmatrix}$. Where it is understood that the right handed singlets do not couple to the $W$'s in the covariant derivatives which are in the $SU(2)$ basis. More generations can easily be added. The left and right handed representations are defined as usual $\Psi _R = \frac{1 + \gamma ^5}{2}\Psi$ and $\Psi _L = \frac{1 - \gamma ^5}{2}\Psi$. So we do not have the terms $i\begin{pmatrix} \bar{\nu}^e _L ,  & \bar{e}_L \end{pmatrix} D_\mu \gamma ^\mu e_R$ ,
$-\begin{pmatrix} \bar{\nu}^e _L , & \bar{e}_L \end{pmatrix}D_5 \gamma ^5 \begin{pmatrix} \nu ^e _L \\ e_L  \end{pmatrix}$, and $-\begin{pmatrix} \bar{\nu}^e _R , & \bar{e}_R \end{pmatrix}D_5 \gamma ^5 \begin{pmatrix} \nu ^e _R \\ e_R \end{pmatrix}$ etc. because all of these terms are zero since $\gamma ^5 \gamma ^\mu = -\gamma ^\mu \gamma ^5$
and $(1 + \gamma ^5 )(1 - \gamma ^5 ) = 0$ where $(\gamma ^5 )^2 = 1$ as usual. \\ With the flux, $\begin{pmatrix} \nu ^e _L \\ e_L \end{pmatrix} \underset{\rm{flux}}{\longrightarrow}\begin{pmatrix} 1 & 0 \\ 0 &  e^{-iby/R}\end{pmatrix}\begin{pmatrix} \nu ^e _L \\ e_L \end{pmatrix}$ (the same for the right handed spinor) and 
$\begin{pmatrix} u _L \\ d_L \end{pmatrix}\underset{\rm{flux}}{\longrightarrow}\begin{pmatrix} e^{\frac{2}{3}iby/R} & 0 \\ 0 & e^{\frac{1}{3}iby/R}\end{pmatrix}\begin{pmatrix} u _L \\ d_L \end{pmatrix}$(and the same for the right handed spinor). Then \\         
$i\begin{pmatrix} \bar{\nu}^e _L ,  & \bar{e}_L \end{pmatrix} D_\mu \gamma ^\mu \begin{pmatrix} \nu ^e _L \\ e_L \end{pmatrix} \underset{\rm{flux}}{\longrightarrow} \\
i\begin{pmatrix} \bar{\nu}^e _L ,  & \bar{e}_L \end{pmatrix}\begin{pmatrix} 1 & 0 \\ 0 & e^{iby/R}\end{pmatrix}
[\partial _\mu + igB W_\mu B^\dagger + \frac{i}{2}g'B_\mu]\begin{pmatrix} 1 & 0 \\ 0 & e^{-iby/R}\end{pmatrix}\gamma ^\mu \begin{pmatrix} \nu ^e _L \\ e_L \end{pmatrix} \\ = i\begin{pmatrix} \bar{\nu}^e _L ,  & \bar{e}_L \end{pmatrix}[\partial _\mu + igW_\mu  + \frac{i}{2}g'B_\mu]\gamma ^\mu \begin{pmatrix} \nu ^e _L \\ e_L \end{pmatrix}$ \\ 
and \\
$-\begin{pmatrix} \bar{\nu}^e _L , & \bar{e}_L \end{pmatrix}D_5 \gamma ^5 \begin{pmatrix} \nu ^e _R \\ e_R \end{pmatrix}\underset{\rm{flux}}{\longrightarrow} \\ -\begin{pmatrix} \bar{\nu}^e _L , & \bar{e}_L \end{pmatrix}\begin{pmatrix} 1 & 0 \\ 0 & e^{iby/R}\end{pmatrix}[\partial _y + igB W_5 B^\dagger + \frac{i}{2}g'B_5]\begin{pmatrix} 1 & 0 \\ 0 & e^{-iby/R}\end{pmatrix}\gamma ^5 \begin{pmatrix} \nu ^e _R \\ e_R \end{pmatrix} \\ = -\begin{pmatrix} \bar{\nu}^e _L , & \bar{e}_L \end{pmatrix}[\partial _y + \begin{pmatrix} 0 & 0 \\ 0 & \frac{-ib}{R}\end{pmatrix} +igW_5 + \frac{i}{2}g'B_5]\gamma ^5 \begin{pmatrix} \nu ^e _R \\ e_R \end{pmatrix}$. Then for the quark terms \, $i\begin{pmatrix} \bar{u} _L ,  & \bar{d}_L \end{pmatrix} D_\mu \gamma ^\mu \begin{pmatrix} u _L \\ d_L \end{pmatrix} \underset{\rm{flux}}{\longrightarrow} \\ i\begin{pmatrix} \bar{u} _L ,  & \bar{d}_L \end{pmatrix}\begin{pmatrix} e^{-\frac{2}{3}iby/R} & 0 \\ 0 & e^{-\frac{1}{3}iby/R}\end{pmatrix}
[\partial _\mu + igB W_\mu B^\dagger + \frac{i}{2}g'B_\mu]\begin{pmatrix} e^{\frac{2}{3}iby/R} & 0 \\ 0 & e^{\frac{1}{3}iby/R}\end{pmatrix}\gamma ^\mu \begin{pmatrix} u _L \\ d_L \end{pmatrix} \\ = 
i\begin{pmatrix} \bar{u} _L ,  & \bar{d}_L \end{pmatrix}
[\partial _\mu + ig\begin{pmatrix} 1 & 0 \\ 0 & e^{-\frac{2}{3}iby/R}\end{pmatrix} W_\mu \begin{pmatrix} 1 & 0 \\ 0 & e^{\frac{2}{3}iby/R}\end{pmatrix}  + \frac{i}{2}g'B_\mu]\gamma ^\mu \begin{pmatrix} u _L \\ d_L \end{pmatrix}$ \\
and \\
$-\begin{pmatrix} \bar{u} _L ,  & \bar{d}_L \end{pmatrix} D_5 \gamma ^5 \begin{pmatrix} u _R \\ d_R \end{pmatrix} \underset{\rm{flux}}{\longrightarrow} \\ i\begin{pmatrix} \bar{u} _L ,  & \bar{d}_L \end{pmatrix}\begin{pmatrix} e^{-\frac{2}{3}iby/R} & 0 \\ 0 & e^{-\frac{1}{3}iby/R}\end{pmatrix}
[\partial _y + igB W_5 B^\dagger + \frac{i}{2}g'B_5]\begin{pmatrix} e^{\frac{2}{3}iby/R} & 0 \\ 0 & e^{\frac{1}{3}iby/R}\end{pmatrix}\gamma ^5 \begin{pmatrix} u _R \\ d_R \end{pmatrix} \\ = 
-\begin{pmatrix} \bar{u} _L ,  & \bar{d}_L \end{pmatrix}[\partial _y + \begin{pmatrix} \frac{2}{3}\frac{ib}{R} & 0 \\ 0 & \frac{1}{3}\frac{ib}{R}\end{pmatrix} + ig\begin{pmatrix} 1 & 0 \\ 0 & e^{-\frac{2}{3}iby/R}\end{pmatrix} W_5 \begin{pmatrix} 1 & 0 \\ 0 & e^{\frac{2}{3}iby/R}\end{pmatrix}  + \frac{i}{2}g'B_5]\gamma ^5 \begin{pmatrix} u _R \\ d_R \end{pmatrix}$. \\ Notice that there are nontrivial couplings for the quark doublet to the $W$'s \\ because of the fluxes. A similar nontrivial coupling does not exist for the electron.
For $-\lambda _e \begin{pmatrix} \bar{\nu}^e _L , & \bar{e}_L \end{pmatrix}\phi e_R$ we have 
$-\lambda _e \begin{pmatrix} \bar{\nu}^e _L , & \bar{e}_L \end{pmatrix}\phi e_R \underset{\rm{flux}}{\longrightarrow} \\
-\lambda _e \begin{pmatrix} \bar{\nu} ^e _L , & \bar{e}_L \end{pmatrix}\begin{pmatrix} 1 & 0 \\ 0 & e^{iby/R}\end{pmatrix}\begin{pmatrix} e^{iby/R} & 0 \\ 0 & 1\end{pmatrix}\phi \, e^{-iby/R}e_R$. Then parameterize $\phi$ as follows, let $\phi  = \frac{1}{\sqrt{2}}\Omega B\begin{pmatrix} 0 \\ v + h\end{pmatrix}$ where $\Omega$ is unitary (remember that $B = \begin{pmatrix} e^{iby/R} & 0 \\ 0 & 1\end{pmatrix}$). Then let $\begin{pmatrix} 1 & 0 \\ 0 & e^{-iby/R}\end{pmatrix}\begin{pmatrix} \nu^e _L  \\ e_L \end{pmatrix}$ transform under the following gauge \\ $ \begin{pmatrix} 1 & 0 \\ 0 & e^{-iby/R}\end{pmatrix}\begin{pmatrix} \nu^e _L  \\ e_L \end{pmatrix}\longrightarrow \Omega \begin{pmatrix} 1 & 0 \\ 0 & e^{-iby/R}\end{pmatrix}\begin{pmatrix} \nu^e _L  \\ e_L \end{pmatrix}$ where now we do not have \\ $\Omega (x^\mu ,y + 2\pi R) = \Omega (x^\mu ,y)$, instead  
\begin{eqnarray}
\phi (x^\mu ,y + 2\pi R) = B(2\pi R)\phi (x^\mu ,y) \nonumber 
\end{eqnarray}
which implies that 
\begin{eqnarray}
\Omega (x^\mu ,y + 2\pi R) B(y + 2\pi R) = B(y + 2\pi R) B^\dagger (y + 2\pi R) \Omega (x^\mu ,y + 2\pi R) B(y + 2\pi R) 
\nonumber
\end{eqnarray}
\begin{eqnarray}
= B(2\pi R)B(y)B^\dagger (y)B^\dagger (2\pi R)\Omega (x^\mu ,y + 2\pi R)B(2\pi R)B(y) \nonumber 
\end{eqnarray}
\begin{eqnarray}
= B(2\pi R)[B^\dagger (2\pi R)\Omega (x^\mu ,y + 2\pi R)B(2\pi R)]B(y) \nonumber
\end{eqnarray}
where $B(y + 2\pi R) = B(2\pi R)B(y) = B(y)B(2\pi R)$ and thus 
\begin{eqnarray}
B^\dagger (2\pi R)\Omega (x^\mu ,y + 2\pi R)B(2\pi R) = \Omega (x^\mu ,y). \nonumber 
\end{eqnarray}
Then \\ 
$-\lambda _e \begin{pmatrix} \bar{\nu} ^e _L , & \bar{e}_L \end{pmatrix}\begin{pmatrix} 1 & 0 \\ 0 & e^{iby/R}\end{pmatrix}\begin{pmatrix} e^{iby/R} & 0 \\ 0 & 1\end{pmatrix}\phi \, e^{-iby/R}e_R \longrightarrow \\ 
-\frac{1}{\sqrt{2}}\lambda _e \begin{pmatrix} \bar{\nu} ^e _L , & \bar{e}_L \end{pmatrix}\begin{pmatrix} 1 & 0 \\ 0 & e^{iby/R}\end{pmatrix}\Omega ^\dagger \Omega \begin{pmatrix} e^{iby/R} & 0 \\ 0 & 1\end{pmatrix}\begin{pmatrix} 0 \\ v + h\end{pmatrix} \, e^{-iby/R}e_R \\ = -\frac{1}{\sqrt{2}}\lambda _e \begin{pmatrix} \bar{\nu} ^e _L , & \bar{e}_L \end{pmatrix}\begin{pmatrix} 0 \\ v + h\end{pmatrix}e_R$ and so 
\begin{equation}
m_e = \lambda _e \sqrt{\frac{v^2}{2}}.
\end{equation}
There is also a flux contribution to this mass and for the quark masses in equations (2) and (3) that follow, but for expediency sake we have included this dependence later in the masses for the modes as the fermion fields require a transformation to include these contributions. See below equation (7).
 
For the quarks, $\begin{pmatrix} e^{\frac{2}{3}iby/R} & 0 \\ 0 & e^{\frac{1}{3}iby/R}\end{pmatrix}\begin{pmatrix} u _L \\ d_L \end{pmatrix} \longrightarrow \Omega \begin{pmatrix} e^{\frac{2}{3}iby/R} & 0 \\ 0 & e^{\frac{1}{3}iby/R}\end{pmatrix}\begin{pmatrix} u _L \\ d_L \end{pmatrix}$ and using the same parameterization for $\phi$ we find \\
$-\lambda _d \begin{pmatrix} \bar{u}_L , & \bar{d}_L \end{pmatrix}\begin{pmatrix} e^{-\frac{2}{3}iby/R} & 0 \\ 0 & e^{-\frac{1}{3}iby/R}\end{pmatrix}\begin{pmatrix} e^{iby/R} & 0 \\ 0 & 1\end{pmatrix}\phi \, e^{\frac{1}{3}iby/R} d_R \longrightarrow \\ -\frac{1}{\sqrt{2}}\lambda _d \begin{pmatrix} \bar{u}_L , & \bar{d}_L \end{pmatrix}\begin{pmatrix} e^{-\frac{2}{3}iby/R} & 0 \\ 0 & e^{-\frac{1}{3}iby/R}\end{pmatrix}\Omega ^\dagger \Omega \begin{pmatrix} e^{iby/R} & 0 \\ 0 & 1\end{pmatrix}\begin{pmatrix} 0 \\ v + h\end{pmatrix} e^{\frac{1}{3}iby/R}d_R \\ = 
-\frac{1}{\sqrt{2}}\lambda _d \begin{pmatrix} \bar{u}_L , & \bar{d}_L \end{pmatrix}\begin{pmatrix} e^{\frac{2}{3}iby/R} & 0 \\ 0 & 1\end{pmatrix}\begin{pmatrix} 0 \\ v + h\end{pmatrix}d_R$ and thus 
\begin{equation}
m_d = \lambda _d \sqrt{\frac{v^2}{2}}.
\end{equation}
Similarly \\ $- \lambda _u \begin{pmatrix} \bar{u}_L , & \bar{d}_L \end{pmatrix}\begin{pmatrix} e^{-\frac{2}{3}iby/R} & 0 \\ 0 & e^{-\frac{1}{3}iby/R}\end{pmatrix}i\tau ^2 \begin{pmatrix} e^{-iby/R} & 0 \\ 0 & 1\end{pmatrix}\phi ^*  \, e^{\frac{2}{3}iby/R}u_R \longrightarrow \\
-\frac{1}{\sqrt{2}}\lambda _u \begin{pmatrix} \bar{u}_L , & \bar{d}_L \end{pmatrix}\begin{pmatrix} e^{-\frac{2}{3}iby/R} & 0 \\ 0 & e^{-\frac{1}{3}iby/R}\end{pmatrix}\Omega ^\dagger i\tau ^2 \Omega ^* \begin{pmatrix} e^{-iby/R} & 0 \\ 0 & 1\end{pmatrix}\begin{pmatrix} 0 \\ v + h\end{pmatrix} e^{\frac{2}{3}iby/R}u_R \\ =
-\frac{1}{\sqrt{2}}\lambda _u \begin{pmatrix} \bar{u}_L , & \bar{d}_L \end{pmatrix}\begin{pmatrix} 1 & 0 \\ 0 & e^{\frac{1}{3}iby/R}\end{pmatrix} \begin{pmatrix} 0 & 1 \\ -1 & 0\end{pmatrix}\begin{pmatrix} e^{-iby/R} & 0 \\ 0 & 1\end{pmatrix}\begin{pmatrix} 0 \\ v + h\end{pmatrix}u_R$ \, 
(where if we let \\ $\Omega = e^{i\vec{\alpha} (x^\mu ,y)\cdot \vec{\tau}}$ then $\tau ^2 \Omega = \Omega ^* \tau ^2$) \, $= -\frac{1}{\sqrt{2}}\lambda _u \begin{pmatrix} \bar{u}_L , & \bar{d}_L \end{pmatrix}\begin{pmatrix} 0 & 1 \\ -e^{-\frac{2}{3}iby/R} & 0\end{pmatrix}\begin{pmatrix} 0 \\ v + h\end{pmatrix}u_R$ which gives 
\begin{equation}
m_u = \lambda _u \sqrt{\frac{v^2}{2}}.
\end{equation} 

Let us now look at the masses for the modes. With the terms \\ $-\begin{pmatrix} \bar{\nu}^e _L , & \bar{e}_L \end{pmatrix}[\partial _y + \begin{pmatrix} 0 & 0 \\ 0 & \frac{-ib}{R}\end{pmatrix}]\gamma ^5 \begin{pmatrix} \nu ^e _R \\ e_R \end{pmatrix}$ and $-\begin{pmatrix} \bar{u} _L ,  & \bar{d}_L \end{pmatrix}[\partial _y + \begin{pmatrix} \frac{2}{3}\frac{ib}{R} & 0 \\ 0 & \frac{1}{3}\frac{ib}{R}\end{pmatrix}]\gamma ^5 \begin{pmatrix} u _R \\ d_R \end{pmatrix}$ and \\ equations (1), (2), and (3) we have 
\begin{equation}
m_{e_n} = \sqrt{\frac{\lambda _e ^2 v^2}{2} + \frac{(n-b)^2}{R^2}}
\end{equation}
\begin{equation}
m_{d_n} = \sqrt{\frac{\lambda _d ^2 v^2}{2} + \frac{(n+\frac{1}{3}b)^2}{R^2}}
\end{equation}
\begin{equation}
m_{u_n} = \sqrt{\frac{\lambda _u ^2 v^2}{2} + \frac{(n+\frac{2}{3}b)^2}{R^2}}
\end{equation}
and
\begin{equation}
m_{{\nu_e}_n} = \frac{|n|}{R}.
\end{equation}
when we integrate out the extra coordinate. Notice that the neutrino remains massless in the zero mode limit as expected. The following transformations were performed in order to get physical mass terms for the fermions, $e_n \longrightarrow e^{i\beta _n \gamma ^5}e_n$ where 
\begin{equation}
e^{2i\beta _n \gamma ^5} = \cos{2\beta _n} + i\gamma ^5 \sin{2\beta _n} = \frac{\lambda _e v}{m_{e_n}} - i\gamma ^5 \frac{n-b}{m_{e_n}R} \nonumber
\end{equation}
as well as  $d_n \longrightarrow e^{i\sigma _n \gamma ^5}d_n$ where 
\begin{equation}
e^{2i\sigma _n \gamma ^5} = \cos{2\sigma _n} + i\gamma ^5 \sin{2\sigma _n} = \frac{\lambda _d v}{m_{d_n}} - i\gamma ^5 \frac{n+\frac{1}{3}b}{m_{d_n}R} \nonumber 
\end{equation}
and $u_n \longrightarrow e^{i\gamma _n \gamma ^5}u_n$ where 
\begin{equation}
e^{2i\gamma _n \gamma ^5} = \cos{2\gamma _n} + i\gamma ^5 \sin{2\gamma _n} = \frac{\lambda _u v}{m_{u_n}} - i\gamma ^5 \frac{n+\frac{2}{3}b}{m_{u_n}R} \nonumber
\end{equation}
and finally $ \nu_{e_n} \longrightarrow i\gamma ^5 \nu_{e_n}$ where again $v = \sqrt{\frac{\mu ^2}{\lambda}}$. 
Now if an orbifold geometry is chosen instead, then the upper half and lower half of the circle are disjoint allowing an extra mathematical degree of freedom for the fields in the model (at the risk of being redundant, I feel it best to be thorough). This extra degree of mathematical freedom is seen by \\ $\phi(x^\mu ,y+\pi R) = \frac{1}{\sqrt{2\pi R}}\sum_{n \, = \, -\infty}^\infty \phi _n (x^\mu)e^{i(n+b)(y+\pi R)/R} = e^{i\pi b}\frac{1}{\sqrt{2\pi R}}\sum_{n \, = \, -\infty}^\infty (-1)^n \phi _n (x^\mu)e^{iny/R}$ where if $n$ is even then the field is even under this inversion of $y$ and if $n$ is odd then the field is odd under this inversion of $y$. 

Therefore we may choose $Z_5$ and $W_5$ to be odd and all of the other fields in this model even. Then once we integrate out the extra coordinate by forming the action of our model, the $W_5$ and $Z_5$ will not couple to any of the other fields and hence, when $b=0$ and $n=0$ for all the fields, gauge invariance is restored as claimed earlier. This also does away with the possibility of the zero mode (standard model) fields coupling to $W_5$ and $Z_5$ which are themselves massive scalar particles in our model. These fields obviously do not exist in the standard model so we needed to make sure they did not couple to any of our "physical" fields.

\end{document}